\documentstyle[aps,pre,floats,graphicx,epsfig]{revtex}
\begin{document}
\draft \twocolumn[\hsize\textwidth\columnwidth\hsize\csname
@twocolumnfalse\endcsname
\title{Critical behavior of the long-range Ising chain from the largest-cluster
probability distribution}
\author{ Katarina Uzelac$^1$, Zvonko Glumac$^2$, Ante Ani\v ci\'c$^1$}
\address{ $^1$Institute of Physics, POB 304, Bijeni\v{c}ka 46, HR-10000 Zagreb, Croatia}
 \address{$^2$Faculty of Electrical Engineering, Kneza Trpimira 2B, 31 000 Osijek, Croatia}
\maketitle

\begin{abstract}

Monte Carlo simulations of the 1D Ising model with ferromagnetic interactions 
decaying with distance $r$ as $1/r^{1+\sigma}$ are performed by applying the 
Swendsen-Wang cluster algorithm with cumulative probabilities.
The critical behavior in the non-classical critical regime corresponding to 
$0.5 <\sigma < 1$ is derived from the finite-size scaling analysis of the 
largest cluster.

\end{abstract}
\vskip2pc]

\narrowtext

The calculation of the critical behavior of the Ising model with long-range (LR)
interactions decaying with distance with a power law $1/r^{d+\sigma}$ is 
not an easy task, even in one dimension, where phase transition at finite 
temperature occurs for $0 < \sigma \leq 1$ \cite{DY69,note1}. 
Exact analytical expressions for the critical exponents may be derived
 for $\sigma \leq 0.5$ \cite{WU82},
corresponding to the classical regime, while for $\sigma > 0.5$ 
only the approximate results exist. Various analytical and numerical 
approaches have been applied, from direct numerical calculations 
on finite chains\cite{NB70}, 
to several approaches based on the renormalization group (RG) and 
scaling\cite{FMN72,Kost76,GU89,MLH90,GU93,CM97}.
The nonlocal character of interactions reduces the efficiency of 
most of the standard approaches so that the values of critical exponents 
obtained by all these methods differ considerably. 

In numerous cases of phase transitions in systems with short-range 
interactions, a useful complementary tool for obtaining both qualitative and 
quantitative results is provided by Monte Carlo (MC) simulations in 
combination with finite-size scaling.

Such systematic studies were lacking for the LR models until recently.
Namely, when applied to models with LR interactions, the standard MC 
approaches based either on Metropolis or on various cluster algorithms 
are particularly time consuming, since the number of operations per 
spin-flip is proportional there to the size of the system. 
Recently, this problem was successfully resolved by Luijten and 
Bl\"ote \cite{LB95} who used the cumulative probabilities within the Wolff 
cluster algorithm\cite{Wolff89}, which they applied to the Ising and similar 
models\cite{LB97,LB96,Luijt99} reducing the computing time by several 
orders of magnitude.

Their very exhaustive studies concentrate on  questions related to the 
mean-field (MF) regime, while very little or no interest has been dedicated 
yet to the regime of the non-classical critical behavior corresponding to 
$\sigma > 0.5$.

The purpose of this work is to extend the MC studies of the critical 
behavior of the LR Ising model to the non-classical regime.
At the same time this is a suitable example to examine the efficiency
of using only the cluster statistics in deriving the critical properties
of the LR model.

The 1D Ising model with LR interactions written in form of a special 
case ($q=2$) of the Potts model is described by the Hamiltonian

\begin{equation}
H = - \sum_{i < j} \; \frac{J}{|i - j|^{1+\sigma}} \; \delta _{s_i, s_j} \; ,
\label{eq:hamilt}
\end{equation}
where $J>0$, $s_i$ is a two-state Potts variable at the site $i$, $\delta$ is the 
Kronecker symbol and the summation is over all pairs of the system.
By the substitution $\delta_{s_i,s_j}=(S_i \cdot S_j +1)/2$, and $J_I=J/2$, where 
$S_i=\pm 1$ and $J_I$ denote the Ising spins and the interaction constant, 
respectively, one recovers the standard definition of the Ising model.
In the mean-field regime, $\sigma \leq 0.5 $, the critical exponents have 
classical values ($\nu=1/\sigma$, $\eta=2-\sigma$).
We focus here on the region $0.5 < \sigma < 1$, where the critical 
exponents are non-trivial and known only approximately. 

When the shortcut using the cumulative probabilities is implemented in
the MC simulations, the full extent of the numerical advantages is achieved 
when only the distribution of spins has to be calculated, while for the energy 
sampling the reduction in the CPU-time is less important.
Recently the calculation by energy sampling was improved\cite{KL00} by
calculating the energy in the momentum space and applying the fast Fourier 
transform. 
Our intent here is to avoid the energy sampling by making a more complete 
study of the cluster statistics.
To this purpose we use the Swendsen-Wang cluster algorithm\cite{SW87}.
The implementation of cumulative probabilities to the Swendsen-Wang 
cluster algorithm is straightforward. 
Each step of this iterative procedure consists in identifying all the clusters 
in a given spin configuration of  the system following the rule that two 
particles belong to the same cluster with the probability $p_{ij}= 
(1 - exp[-J(i,j)/T])\delta_{s_i,s_j}$, and then flipping all the clusters randomly.
The cumulative probabilities are applied at the point of identifying the 
individual clusters along the same lines as was done\cite{LB95} for the 
Wolf single-cluster algorithm, and this reduces the number of required 
operations per a single spin-flip by a factor of system size.
Like the Wolf algorithm, the Swendsen-Wang algorithm suppresses the 
slowing-down at criticality, and though it might be somewhat more costly 
regarding CPU time, it gives a more complete insight into the cluster statistics 
and the related probability distributions, such as the distribution of cluster 
sizes or that of the largest cluster.

The probability distributions related to clusters are mostly used in the 
description of critical properties of geometrical transitions, like the 
percolation\cite{stauffer}. 
As it follows from the graph expansion by Kasteleyn and Fortuin
\cite{KF69}, the very basis of the Swendsen-Wang algorithm, this cluster 
statistics may be related to the thermodynamic quantities of the thermal 
transitions as well. 
The only important difference is that in percolation one deals with
the simple geometrical clusters, while in thermal transitions the clusters 
formed by the {\it active bonds} (in terms of the cluster algorithm) 
are considered. In systems with LR interactions, such as the present model, 
this difference is obvious.

The cluster-related distributions were much less exploited in the 
case of thermal transitions, although some detailed studies exist 
(see e.g. refs \cite{CK81,OMHB90}) for models with short-range interactions.
More recently, there has been renewed interest in this subject within a 
somewhat different context \cite{MCLSC95-6,Bazant00}.

By performing $10^5$ MC iterations, which is sufficient for most of data 
presented here, the chains of sizes up to a few tens of thousands of sites 
could be simulated without requiring too much numerical effort on a modest 
workstation.

We limit our analysis here to the size distribution of the largest cluster, the 
quantity that was left out of reach within the earlier single-cluster approach.

The probability that the largest cluster will be of the size $l$ is  defined as
\begin{equation}
      P_{max}(l)  = \; \frac{1}{n_{MCS}} \  N_{max}(l),
\label{eq:prob}
\end{equation}
where $N_{max}(l)$ is the total number of occurrences of the largest 
cluster of  size $l$ during $n_{MCS}$  MC swaps of the system.

Several thermodynamic quantities of interest may be expressed through the 
corresponding moments defined as 
\begin{equation}
     <l^k> = \sum_l l^k P_{max}(l).
\label{moms}
\end{equation} 

The first moment 
\begin{equation}
     <l> = \sum_l l P_{max}(l)  
\label{mom1}
\end{equation}
gives the mean size of the largest cluster. Below the critical temperature 
it describes the order parameter, which is usually defined as the average 
of the largest component, i.e. $M = (q <max\{m_\alpha\}>-1)/(q-1)$, where 
$m_\alpha$ denotes the fraction of spins of the system in the state $\alpha$.

A quantity of interest to consider is the Binder's fourth-order cumulant
\cite{BIN81} ratio for the largest cluster,
\begin{equation}
     R_L(T) =    \; \frac{<l^4>}{<l^2>^2},
\label{mom4}
\end{equation}
defined in the same way as for the standard order parameter.
Differences between the two corresponding distribution functions,
however, result in different shapes of the same ratio defined for
the two quantities.
As may be seen in Fig. 1,  the ratio $R_L(T)$ at high temperatures 
and in the thermodynamic limit tends to 1 and not to 3, while
the corresponding probability distribution is expected to be
non-Gaussian, as found already in the case of percolation 
\cite{MH84} and the short-range Ising model \cite{OMHB90}.
However, as well as in the mentioned cases,  these differences 
should not change the basic scaling properties in the critical regime.
At low temperatures and around $T_c$ the behavior of the two 
ratios coincides.
Like in the case of the standard order parameter, there is a common crossing 
point of the curves $R_L(T)$ of different sizes where the ratio (\ref{mom4}) 
almost does not depend on $L$ and  which can be identified as the 
critical temperature.
In Fig. 1, $R_L(T)$ is shown in the case $\sigma=0.9$ and for sizes 
varying from 1000 to 20000. 
\begin{figure}[h!!!]
\begin{center}
\leavevmode
\epsfig{file = 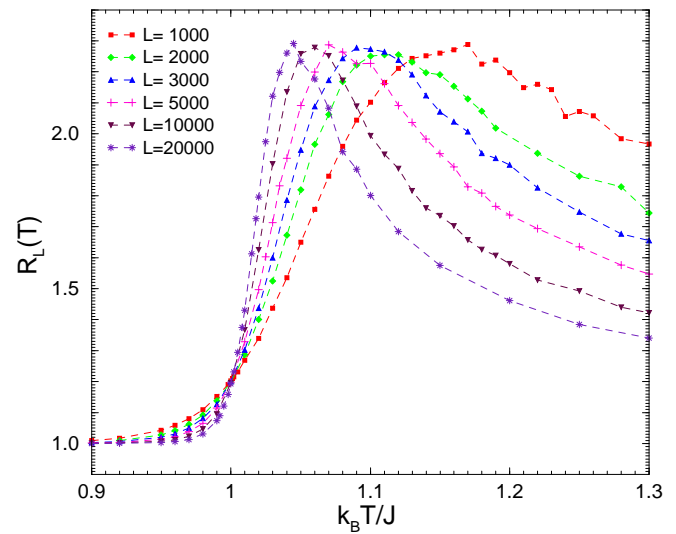, width = 1. \linewidth}
\caption[fig1]{
Ratio $R_L(T)$ for $\sigma = 0.9$ and sizes L varying from 1000 to 20000.
}
\label{fig1}
\end{center}
\end{figure}
By taking into account sizes within the range $1000 \leq L \leq 20000$, the 
determination of this crossing point can be done with precision of 4 digits, 
which gives the estimated error of $0.2\%$.
\begin{table}[[h!!!]
\caption [ ]{
Comparison of results for the critical temperature $T_c$ to earlier results by 
Nagle and Bonner\cite{NB70}, Monroe et al.\cite{MLH90},  FRS\cite{GU93},
Cannas and Magalh\~aes \cite{CM97} and Monroe\cite{M99}.
}
\label{tab1}
\begin{center}
\begin{tabular}{llllllll}
$\sigma$ & this work & \cite{NB70}& \cite{MLH90} & \cite{GU93} &  \cite{CM97} & \cite{M99} \\
\hline
 0.6	& $1.770\pm .003$    &1.766     &  1.7885  & 1.7718   & -        &	-        \\
 0.7    & $1.463\pm .003$    &1.458     &  1.491    & 1.4635   & 1.28    &  1.466850 \\
 0.8    & $1.2155\pm .002$  &1.212     &  1.2585   & 1.2150   & -        &	-          \\
 0.9	& $1.001  \pm .001$  & 1.0015  & 1.058      & 1.0027 & 0.77 & 1.018845 \\
\end{tabular}
\end{center}
\end{table}

In table I are presented the critical temperatures for several values of 
$\sigma$ in the non-classical regime $0.5<\sigma<1$ incremented by 
$0.1$, obtained as common crossings of $R_L(T)$. 
For comparison, we quote the earlier results obtained by  
numerical calculations \cite{NB70}, coherent anomaly approach
\cite{MLH90,M99}, finite-range scaling (FRS) \cite{GU93}, and 
real-space RG \cite{CM97}.
When quoting ref. \cite{MLH90} we mention only one of the two given 
sets of results for $T_c$, which fits better to the later 
work \cite{M99} performed with a significantly improved precision, but, 
unfortunately, available only for two values of the range considered here.

Near $T_c$ (i.e. in the regime $L < \xi$), the moments (\ref{moms}) and 
their ratio (\ref{mom4}) should obey finite-size scaling. Thus 
\begin{equation}
R_L(T) = L^{x} \  f (L^{1/\nu} \ \tau) ,  
\label{scal}
\end{equation}
where $\tau =  (T - T_c)/{T_c}$ is the reduced temperature. 
Since $R_{\infty}(T_c)$ is finite, $x=0$.

It turns out that the critical exponent $\nu$ may be estimated with a rather 
good precision by a simple tuning of the unknown exponent $1/\nu$ in the 
Eq.  (\ref{scal}) until the curves collapse in the vicinity of the crossing point. 

In Fig. 2 we present the curves from Fig. 1, collapsed after taking $1/\nu = 0.40$.
\begin{figure}[h!!!]
\begin{center}
\leavevmode
\epsfig{file = 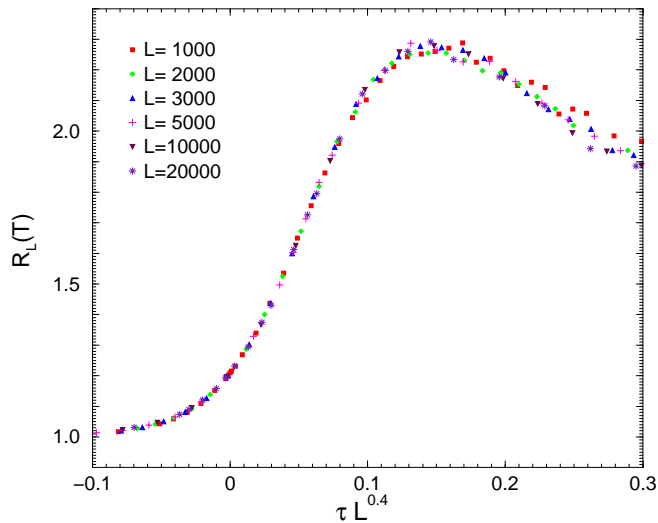, width = 1. \linewidth}
\caption[fig2]{
Ratio $R_L(T)$ for $\sigma = 0.9$ 
versus reduced temperature $\tau$ scaled with $L^{1/\nu}$}
\label{fig2}
\end{center}
\end{figure}

The results for $\nu^{-1}$  obtained by collapsing fits for other considered 
values in the interval $0.5 < \sigma < 1$ are given in table  \ref{tab2}.
\begin{table}[[h!!!]
\caption [ ] {Results for $\nu^{-1}$ compared with earlier results of Nagle and 
Bonner\cite{NB70}, Fisher et al.\cite{FMN72}, Monroe et al. \cite{MLH90}, 
FRS\cite{GU93} and Cannas and Magalaes \cite{CM97}. 
For the meaning of labels a and b, see the text.} 
\label{tab2}
\begin{center}
\begin{tabular}{llllllll}
$\sigma$   &  this work   &  \cite{NB70}a ~\cite{NB70}b &  \cite{FMN72}  & \cite{MLH90}a ~ \cite{MLH90}b   & \cite{GU93} &\cite{CM97}  \\
\hline
0.6  & $0.50\pm0.02$ &  0.504  0.540 &  0.5098 &  0.441  0.471  & 0.501 &  ~~- \\
0.7  & $0.51\pm0.02$ &  0.542  0.552 &  0.4856 &  0.478  0.461 &  0.518 &  0.376 \\
0.8  & $0.47\pm0.02$ &  0.533  0.538 &  0.4461 &  0.505  0.446 &  0.483 &  ~~- \\
0.9  & $0.40\pm0.01$ &  0.529  0.526 &  0.4032 &  0.542  0.421 &  0.405 &  0.256\\
\end{tabular}
\end{center}
\end{table}
In order to reduce the degree of arbitrariness of such fits we require the curves 
to be indistinguishable in the region of temperatures around $T_c$, up to the 
temperature of the maximum of $R_L(T)$, $T_{max}$. This temperature
may be related to the finite-size crossover temperature up to which 
the finite-size scaling equation (\ref{scal}) should hold.
In Fig. \ref{fig1} one can observe that the maximum of  $R_L(T)$ shrinks
and shifts towards $T_c$ with increasing L.
The rough fit of the difference  $T_{max} -T_c$ to the power-law form $(1/L)^{\phi}$  
gives the values of $\phi$ which roughly correspond to $1/ \nu$ for a given 
$\sigma$, as expected for the finite-size crossover exponent \cite{FB72}.
The interval of tuning in which the curves remain indistinguishable 
determines the error margins of the obtained values for the exponent 
$\nu^{-1}$, which increase with decreasing $\sigma$ but do not exceed  
4\% in the considered range of $\sigma$'s. 

Alternative calculations of the exponent, e.g. by using the derivatives of 
moments, turn out to be less advantageous. 
(The purely numerical derivative would require a much higher numerical 
precision and consequently far more extensive runs;
the derivative obtained by histogram interpolations or by mixed moments
would require an energy sampling and an increased numerical precision 
in addition.)
We also mention at this point that the calculations of the Binder's cumulant 
ratio may equally be performed for the standard order parameter using the 
Wolf cluster algorithm. 
One advantage of the approach which uses the Swendsen-Wang cluster 
algorithm is that it has similar precision and efficiency both in low- and 
high-temperature regimes, which improves the scaling fit.

The degree of precision of the obtained data permits the comparison to the 
results by other approximate methods.
These results, which are quite few, are quoted in table \ref{tab2} in chronological 
order and correspond to numerical calculations on finite chains using 
Pad\'e series extrapolations \cite{NB70}, $\epsilon$-expansion
near $2\sigma=d$ \cite{FMN72}, the coherent anomaly method \cite{MLH90}, 
finite-range scaling \cite{GU93}, and real-space RG \cite{CM97}.
In refs \cite{NB70} and \cite{MLH90} only the numerical data for the 
exponents $\gamma$ and $\beta$ are available. 
The exponent $\nu$ may be expressed in two ways, which, due to the 
approximate character of the results, do not give identical values.
By "a" are labelled the values obtained from the scaling relation 
$\nu = \gamma/(2-\eta)$ and the exact analytical expression $\eta=2-\sigma$ 
expected to hold also in the whole non-classical regime.
By "b" are labelled the values obtained from the scaling relation involving 
both exponents, $\nu = \gamma + 2 \beta$.
The quoted results of ref. \cite{FMN72} are the ones obtained by the expansion
in $\Delta\sigma = \sigma-0.5$ up to the second order.
The expansion in $(1-\sigma)$ \cite{Kost76} is not quoted in the table. 
It holds very close to $\sigma= 1$ and we may quote only the value for 
$\sigma=0.9$ which is equal to $\nu^{-1} =0.447$.
The results of ref.  \cite{FMN72} are also expected to be more reliable closer 
to $\sigma=0.5$.
Other approaches cover the whole range of $\sigma$ evenly,
but their results are rather different and their accuracy as well.
Our MC results agree the best with the FRS results. 

In  ref. \cite{CM97}, which deals with the Potts model with arbitrary number 
of states q, we found intriguing the conjecture that the exponent $\nu$ would 
remain the same for other values of $q$ (when the transition is of the second-order),
which is in contrast to earlier results of FRS \cite{GU93}.
A MC approach can thus be used to provide an independent calculation. 
(It also has an advantage over both used approaches that it is able to distinguish 
the first- from the second-order transition\cite{GU98,KL00}). 
Our first calculations \cite{UGprep}, performed along the same lines as above, 
give for the case $q=3$, $\sigma=0.9$ the value $1/\nu=0.48\pm0.01$, 
which would rather be in favor of the FRS results.

In conclusion, we have applied the MC simulations to the two-state Potts 
model with LR interactions on chains up to 20000 sites by using the 
Swendsen-Wang algorithm to calculate the cluster statistics, in particular the
size distribution of the largest cluster. 
The approach was used as an alternative calculation of the critical 
behavior in this model in the non-classical regime $0.5<\sigma<1$. 
We have shown that the scaling analysis of the Binders 4th-order cumulant 
ratio of the largest-cluster size gives the critical temperature and 
the critical exponent $\nu$ with reasonable accuracy (better then 0.2\% 
and 4\%, respectively).
Although the overlapping fit is not expected to be very precise, 
nor were the simulations pushed to their extreme, the estimated 
errors are significantly smaller than the differences between the results 
obtained by different approaches. 
The obtained results, presented in tables I and II, are closest to the values 
obtained earlier by the FRS. 
While the present work deals only with the Binder's 4th-order cumulant ratio, 
a wider analysis, which includes other quantities that may be derived from the 
cluster statistics, should be performed in future.
We believe that the approach through cluster statistics might be useful to 
study other cases of the Potts model and other models with discrete 
symmetries, where in the case of LR interactions, the alternative methods 
are very restricted.



\begin{thebibliography}{...}
\bibitem{DY69} F. J. Dyson, Commun. Math. Phys. {\bf 12}, 91 (1969).
\bibitem{note1} Rigurous proof for the transition at finite $T_c$ in the most 
important case $\sigma=1$, characterized by a defect-mediated critical 
behavior and related to the Kondo problem (P.W. Anderson and  G.Yuval, 
Phys.Rev. {\bf B1}, 1522 (1970)) was given much later (J.Fr\"ohlich and T. 
Spencer, Comm. Math. Phys. {\bf 84},87 (1982) ), much after Thouless result 
(D.J. Thouless, Phys.Rev. {\bf 187}, 732 (1969) 
on the discontinuous nature of the order parameter in this case.
\bibitem{WU82}  F. Y. Wu,  Rev. Mod. Phys. {\bf 54}, 235 (1982).
\bibitem{NB70} J. F. Nagle and J. C. Bonner,  J. Phys. C {\bf 3}, 352 (1970).
\bibitem{FMN72} M. E. Fisher, S. K. Ma and B. G. Nickel, Phys. Rev. Lett. {\bf 29}, 917 (1972).
\bibitem{Kost76} J. M. Kosterlitz, Phys. Rev. Lett. {\bf 37}, 1577 (1976).
\bibitem{GU89}  Z. Glumac and K. Uzelac, J. Phys. A {\bf 22}, 4439 (1989).
\bibitem{MLH90} J. L. Monroe, R. Lucente and J. P. Hourlland, J. Phys. A {\bf 23}, 2555 (1990).
\bibitem{GU93}  Z. Glumac and K. Uzelac, J. Phys. A {\bf 26}, 5267 (1993).
\bibitem{CM97}  S. A. Cannas  and A. C. N. de Magalh\~aes,  J. Phys. A {\bf 30}, 3345 (1997).
\bibitem{LB95} E. Luijten and H. W. J. Bl\"ote, Int. J. Mod. Phys. C {\bf 6}, 359 (1995).
\bibitem{Wolff89} U. Wolff, Phys. Rev. Lett. {\bf 62}, 361 (1989).
\bibitem{LB97} E. Luijten and H. W. J. Bl\"ote, Phys. Rev. B {\bf 56}, 8945 (1997).
\bibitem{LB96}  E. Luijten and H. W. J. Bl\"ote, Phys. Rev. Lett. {\bf 76}, 1557 (1996).
\bibitem{Luijt99} E. Luijten, Phys. Rev. E {\bf 60}, 7558 (1999). 
\bibitem{KL00} M. Kretch and E. Luijten, Phys. Rev. E {\bf 61}, 2058 (2000).
\bibitem{SW87} R. H. Swendsen and J.-S. Wang, Phys. Rev. Lett. {\bf 58}, 86 (1987).
\bibitem{stauffer} D. Stauffer, Phys. Rep. {\bf 54}, 1 (1979).
\bibitem{KF69} P. W. Kasteleyn and C. M. Fortuin, J. Phys. Soc. Jpn.
Suppl. {\bf 26} , 11 (1969); C. M. Fortuin and P. W. Kasteleyn, Physica {\bf 57}, 536 (1972).
\bibitem{CK81} A. Coniglio and  W. Klein, J. Phys. A {\bf 13}, 2775 (1980).
\bibitem{OMHB90} M. D'Onorio de Meo, D. W. Heermann, K. Binder, J. Stat. Phys. {\bf 60}, 
585 (1990).
\bibitem{MCLSC95-6} J. Machta, Y. S. Choi, A. Lucke, T. Schweizer and L. M. Chayes, 
Phys. Rev. Lett. {\bf 75}, 2492 (199e); Phys. Rev. E {\bf 54}, 1332 (1996).
\bibitem{Bazant00} M. Z. Bazant, Phys. Rev. E {\bf 62}, 1660 (2000).
\bibitem{BIN81} K. Binder, Phys. Rev. Lett. {\bf 47}, 693 (1981).
\bibitem{MH84} A. Margolina and H.J. Herrmann, Phys. Lett. {\bf 104A}, 295 (1984).
\bibitem{M99} J. L. Monroe, J. Phys. A {\bf 32}, 7083 (1999).
\bibitem{FB72} M. E. Fisher and M. N. Barber,  Phys. Rev. Lett. {\bf 28}, 1516 (1972).
\bibitem{GU98} Z. Glumac and K. Uzelac, Phys. Rev. E {\bf 58}, 4372 (1998).
\bibitem{UGprep} K. Uzelac and Z. Glumac, in preparation 
\end{thebibliography}
\end{document}